\documentclass[aps,prl,twocolumn,superscriptaddress,calc,10pt,floatfix]{revtex4-2}

\usepackage{graphicx}
\usepackage{amssymb,amsmath}
\usepackage{wasysym}
\usepackage{pifont}
\usepackage{float}
\usepackage{physics}
\usepackage{siunitx}
\usepackage[dvipsnames]{xcolor}
\usepackage[normalem]{ulem} 
\usepackage[pdftex,breaklinks,colorlinks=true,linkcolor=blue,citecolor=blue,urlcolor=blue]{hyperref}
\usepackage{tikz}
\usetikzlibrary{calc,shapes}

\newcommand{\mitCUAaddress}{Department of Physics, MIT-Harvard Center for Ultracold Atoms, and Research Laboratory of Electronics, MIT, Cambridge, Massachusetts 02139, USA}

\begin{document}

\thickmuskip = 3mu

\title{Competition of fermion pairing, magnetism, and charge order in the spin-doped attractive Hubbard gas}
\date{\today}
\author{Thomas Hartke}
\thanks{These authors contributed equally.}
\affiliation{\mitCUAaddress}

\author{Botond Oreg}
\thanks{These authors contributed equally.}
\affiliation{\mitCUAaddress}

\author{Chunhan Feng}
\affiliation{Center for Computational Quantum Physics, Flatiron Institute, 162 5th Avenue, New York, New York 10010, USA}

\author{Carter Turnbaugh}
\affiliation{\mitCUAaddress}

\author{Jens Hertkorn}
\affiliation{\mitCUAaddress}

\author{Yuan-Yao He}
\affiliation{Institute of Modern Physics, Northwest University, Xi'an 710127, China}

\author{Ningyuan Jia}
\affiliation{\mitCUAaddress}

\author{Ehsan Khatami}
\affiliation{Department of Physics and Astronomy, San Jos\'e State University, San Jos\'e, CA 95192, USA}

\author{Shiwei Zhang}
\affiliation{Center for Computational Quantum Physics, Flatiron Institute, 162 5th Avenue, New York, New York 10010, USA}

\author{Martin Zwierlein}
\affiliation{\mitCUAaddress}

\begin{abstract}
The tension between fermion pairing and magnetism affects numerous strongly correlated electron systems, from high-temperature cuprates to twisted bilayer graphene. Exotic forms of fermion pairing and superfluidity are predicted when attraction between fermions competes with spin doping. 
Here, we follow the evolution of fermion pairing and charge and spin order in a spin-imbalanced attractive Hubbard gas of fermionic $^{40}$K atoms, covering a wide range of densities, magnetizations, and interactions with single-atom 
resolution.
At low spin imbalance and weak interactions, we find a mixture of 
nonlocal fermion pairs coexisting with itinerant excess fermions.
For stronger interactions an effective hard-core Bose-Fermi mixture emerges.
Spin doping drives a crossover from charge-density wave correlations to a Fermi liquid of polarons.
Beyond a certain spin imbalance and interaction strength, we find evidence for the onset of combined spin- and pair-density wave order, a possible precursor for the existence of magnetized superfluidity in the attractive Hubbard system.
\end{abstract}

\maketitle


Fermionic superfluidity and magnetism appear, at first sight, to be mutually exclusive. Superfluidity of $s$- or $d$-wave character requires pairing of fermions of opposite spin while a magnetic field forces spins to align. However, 
exotic phases such as the Fulde-Ferrell-Larkin-Ovchinnikov (FFLO)-phase are predicted to feature a magnetized superfluid where excess spins order in the presence of a spatially modulated pair condensate~\cite{fulde1964superconductivity, larkin1965nonuniform}.
Spin-imbalanced superfluidity has been investigated with atomic Fermi gases~\cite{Radzihovsky_2010, Zwierlein2014}, revealing the Pauli limit of superfluidity with increasing spin imbalance~\cite{Zwierlein_2006}. However, for 3D bulk gases, the parameter regime where exotic phases are predicted is narrow, and phase separation between a fully paired region and a magnetized normal state prevails~\cite{Shin2008}.

\begin{figure}[!tbp]
	\centering
	\includegraphics[width=\columnwidth]{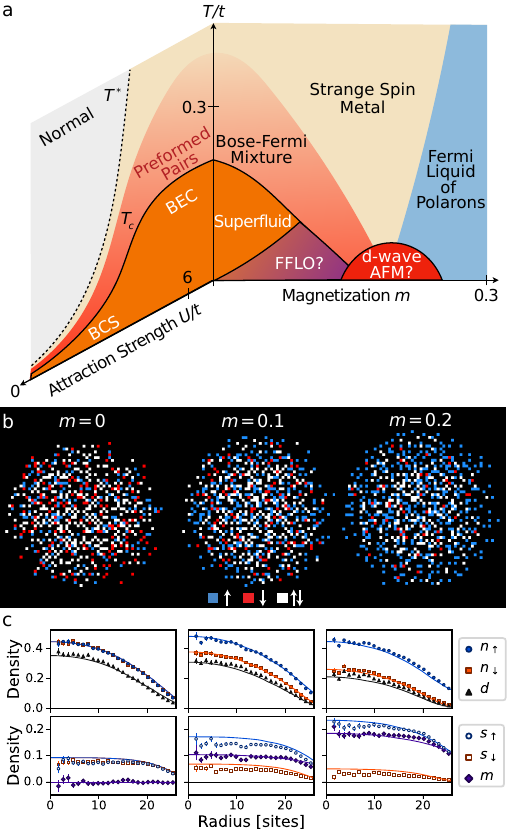}
	\caption{\textbf{Realizing the spin-doped attractive Hubbard model.} 
        {\bf (a)}~Schematic phase diagram of the spin-doped attractive Hubbard model away from half filling (density $n<1$) at magnetization $m$. The spin-balanced gas features a BEC-BCS crossover below a critical temperature $T_C$ and preformed pairs above $T_C$ that display charge-density wave (CDW) and superfluid (SF) correlations. With increasing magnetization, exotic forms of superfluidity (FFLO) are predicted, including strange spin metal behavior at elevated temperatures and a Fermi liquid of polarons at high polarization.
        {\bf (b)}~Spin- and atom-resolved images of a trapped spin-doped Hubbard gas with central magnetization $m = 0$, $m = 0.1$, and $m = 0.2$ at interaction strength $U/t=5.8(3)$, detected via quantum gas microscopy. 
        {\bf (c)}~Upper row:~radially averaged densities of spin-up~($n_\uparrow$) atoms, spin-down~($n_\downarrow$) atoms, and doubly occupied sites~(doublons, $d$) as a function of radial position in the trap.
        Lower row:~singly occupied site (singlon) densities, where $s_\uparrow = n_\uparrow - d$, $s_\downarrow = n_\downarrow - d$, and magnetization $m=n_\uparrow - n_\downarrow$. 
        The flat dependence of magnetization on position in the trap is indicative of a Bose-Fermi mixture where excess fermions are repelled by bosonic pairs that compensate the trapping potential. Solid lines are from AFQMC calculations (see main text).
        }
	\label{fig:IntroFigure}
\end{figure}

The attractive Hubbard model on a 2D square lattice provides a rich setting to study the competition between fermion pairing, charge order, and spin order. Proposed to describe bismuthate superconductors~\cite{Micnas1990Superconductivity, Auerbach1994}, the model is believed to host a large variety of phases, including exotic superfluid states of FFLO type~\cite{Moreo2007, Loh2010FFLO}. Remarkably, the spin-doped attractive model maps exactly to the elusive charge-doped repulsive Hubbard model~\cite{Shiba1972, Emery1976, Moreo2007, ho2009quantum, chiesa2013phases}, potentially holding the key to high-temperature superconductivity in the cuprates. The $XY$-, and $Z$-antiferromagnets of the repulsive model transform onto the superfluid and charge-density wave ordered states in the attractive model. Similarly, stripe ordered phases map onto the FFLO state and spin-charge-density waves.
Despite decades of intense effort, exact analytic or numerical solutions are not available except for special parameter regimes, restricted lattice sizes, or particular topologies. Conventional quantum Monte-Carlo calculations of these models suffer from the fermion sign problem~\cite{loh1990sign}.


Figure~\ref{fig:IntroFigure}(a) gives an overview of the various expected phases of the attractive Hubbard model.
The spin-balanced case away from half-filling features a crossover from Bardeen-Cooper-Schrieffer (BCS) superfluidity at weak attraction to Bose-Einstein condensation (BEC) of small pairs at strong interaction~\cite{pincus1973correlated, scalettar1989phase}. The presence of non-zero magnetization $m = n_\uparrow - n_\downarrow$
gives rise to a crucial tension between full pairing of minority fermions, with excess majority fermions not profiting from the pairing gap, and a complete quenching of pairing, with the emergence of a spin-imbalanced Fermi mixture~\cite{Radzihovsky_2010,Zwierlein2014}.
For small magnetization, the FFLO state is expected to appear~\cite{Moreo2007, Loh2010FFLO}. At low temperatures and intermediate magnetization, a putative $d$-wave anti-ferromagnetic state would be the counterpart of the elusive $d$-wave superconducting state in the repulsive system.
Above the critical temperature for superfluidity, one expects a fluctuating regime of preformed pairs coexisting with excess fermions, featuring superfluid and charge-density wave correlations.
In the limit of large imbalance and sufficient filling fraction, calculations show that even for large attraction, a single minority spin remains unpaired, forming a dressed quasiparticle: the Fermi polaron~\cite{Sorella1994, Chevy2006, Prokofev2008a, Schirotzek2009, kohstall2012metastability, ness2020observation, Pascual2024, Massignan2025}.
The ground state at large imbalance is thus expected to be a Fermi liquid of polarons.
At intermediate magnetizations, in analogy with the strange metal phase of the cuprates, one may expect a regime featuring anomalous spin transport, a ``strange spin metal.''

To investigate fermion pairing, charge order, and spin order throughout these exotic regimes, we realize the spin-doped attractive Hubbard model using a Fermi gas of $^{40}$K atoms in a single 2D plane of an optical lattice, employing full spin- and charge-resolved imaging at single atom resolution~\cite{hartke2020doublon,koepsell2020robust,Hartke2022Direct}. Interaction strength and magnetization can be freely controlled. An underlying trapping potential varies the density from near-half filling in the center to a dilute spin mixture in the wings of the cloud. 
The temperature $T/t\approx 0.3$, as measured from density fluctuations~\cite{hartke2020doublon}, is above the predicted superfluid critical temperature~\cite{paiva2004critical, He2022}, but well in the pseudo-gap regime of preformed pairs~\cite{Hartke2022Direct}, where precursors to the lower temperature states can be observed.
We also perform systematic computations using state-of-the-art auxiliary-field quantum Monte Carlo (AFQMC)~\cite{Zhang1999, He2019, BSS} and numerical linked-cluster expansion (NLCE)~\cite{M_rigol_06, b_tang_13b} to make direct comparisons with experiment.

Figure~\ref{fig:IntroFigure}(b) shows single experimental snapshots of the atom distribution for various degrees of spin-doping at attraction strength $U = 5.8(3) t$, in the strongly correlated regime of the Hubbard model. Radial averages of the densities of each spin species, the local magnetization, and the densities of singly and doubly-occupied sites (singlons and doublons) 
along with AFQMC results are reported in Fig.~\ref{fig:IntroFigure}(c).
For zero magnetization ($m=0$) the number of doublons $d$ is strongly enhanced compared to a noninteracting system, for which $d$ would just be given by the product of spin densities, $n_\uparrow n_\downarrow$. All fermions are paired, 
but pairing is nonlocal causing the appearance of singlons~\cite{Hartke2022Direct}.

With spin doping, not every majority spin can be part of a pair.
This tension is apparent in the measured singlon densities (lower panels of Fig.~\ref{fig:IntroFigure}c). For simple two-body pairing the nonlocal fraction, imaged as singlons, should be in fixed proportion to the local part, the doublon fraction. Instead, the singlon density is nearly homogeneous while the doublon density varies with position in the trapping potential.
Consequently, the magnetization $m = n_\uparrow - n_\downarrow = s_\uparrow - s_\downarrow$ also displays a nearly flat profile (Fig.~\ref{fig:IntroFigure}(c), middle and right panel). This near-homogeneous magnetization is in stark contrast to the situation of spin-imbalanced Fermi gases in the continuum: for two and three dimensions, a central spin-balanced fully paired core strongly repels a surrounding spin-imbalanced Fermi liquid~\cite{Zwierlein_2006,Shin2006Observation,Mitra2016Phase}, while in resonant 1D Fermi gases, excess fermions interact weakly with pairs and reside in the trap center~\cite{Liao2010}. Instead, the present case of the 2D attractive Hubbard gas lies in between these two extremes.
The uniform magnetization indicates coexistence of excess fermions and bosonic pairs, and implies that their mutual repulsion is compensating the attraction from the trap.
Such homogeneous magnetization should enhance prospects for FFLO-type superfluidity~\cite{he2006finite,koponen2007finite}.
The magnetization only decreases in the outer wings of the trapped gas, which is the characteristic behavior of an unpaired, spin-imbalanced Fermi mixture~\cite{Zwierlein_2006}.


\begin{figure}[!t]
	\centering
	\includegraphics[width=\columnwidth]{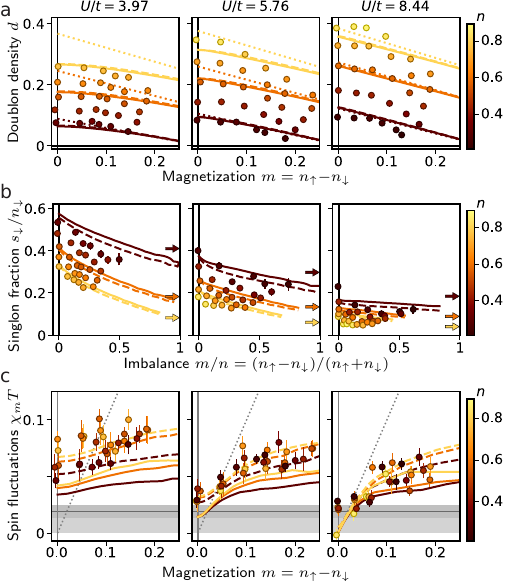}
	\caption{\textbf{Evolution from fermion pairs to a Fermi liquid of polarons.}
{\bf (a)}~The doublon density $d$, measuring the strength of short-range pair correlations, evolves smoothly from the regime of paired fermions at spin balance to the expected polaronic Fermi liquid at large magnetization. Dotted lines give the doublon density for a gas of $n_\downarrow = (n-m)/2$ Fermi polarons at $T=0$ from the variational Ansatz~\cite{Chevy2006, Pascual2024, SI}.
{\bf (b)}~The fraction of singlons among minority atoms $s_\downarrow/n_\downarrow$ quantifies nonlocal pair correlations, and gradually decreases with $m$ to the limit expected for Fermi polarons~\cite{Pascual2024, SI} (arrows).
{\bf (c)}~The vanishing of total magnetization fluctuations $\sum_\delta \langle \hat{m}_i \hat{m}_{i+\delta}\rangle_c = \chi_m T$ indicates full fermion pairing at $m = 0$ beyond $U/t \gtrsim 5.76$. Fluctuations increase with magnetization but remain sub-poissonian (dotted line is $\chi_m T = m$), implying a degenerate Fermi gas of excess fermions at small $m$, and a degenerate Fermi mixture at large $m$.
The gray shaded region is the noise floor for detecting fluctuations, given by the standard deviation at $m=0$ and $U/t=11.93$. 
Lines are AFQMC results at $T/t=0.33$~(solid) and $T/t=0.50$~(dashed).}
	\label{fig:PairingFigure}
\end{figure}

\begin{figure*}[!t]
	\centering
	\includegraphics[width=7 in]{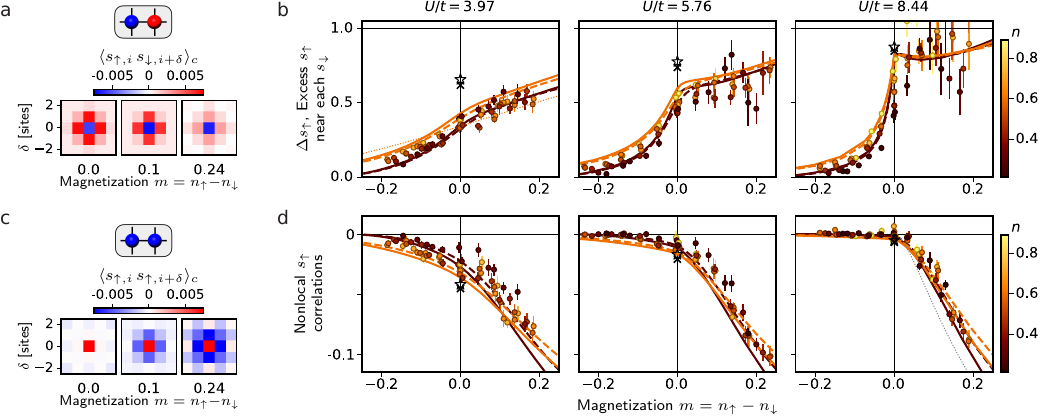}
	\caption{\textbf{
Spatial structure of pair correlations.
 }
{\bf (a)}~Inter-spin connected singlon correlations $\langle s_{\uparrow,i} s_{\downarrow, i+\delta} \rangle_c$ at $U/t=5.76$ and $n\approx 0.7$. Bunching is observed between spin-up and spin-down singlons at all magnetizations, whether due to pairing or polaronic attraction. 
{\bf (b)}~Excess density of spin-up singlons surrounding an isolated spin down ($\Delta s_\uparrow = \sum_{\delta \neq 0}\langle s_{\uparrow,i} s_{\downarrow, i+\delta} \rangle_c/s_\downarrow$) versus magnetization, at various densities and interaction strengths. For two-body pairing each isolated spin is located near exactly one opposite spin, thus $\Delta s_\uparrow =1$. An excess density less than one in a gas with full pairing (at $U/t = 8.44$ and $5.76$ and $m \lesssim 0.05$) implies that each spin attracts only a portion of an opposite spin, a many-body form of pairing similar to a BCS gas.
{\bf (c)}~Intra-spin connected correlations $\langle s_{\uparrow,i} s_{\uparrow, i+\delta} \rangle_c$ between isolated alike fermions at $U/t=5.76$ and $n\approx 0.7$.
At large magnetization, negative correlations develop: the Fermi hole forms due to Pauli exclusion.
{\bf (d)}~Nonlocal correlations ($\sum_{\delta \neq 0} \langle s_{\uparrow,i} s_{\uparrow, i+\delta} \rangle_c$) of majority singlons ($s_\uparrow$ for $m > 0$) converge at large $U/t$ to the expectation for a degenerate Fermi gas of excess spins (gray dotted line, right panel: free Fermi gas of dopants at $T/t = 0.33$). Minority singlons ($s_\uparrow$ for $m < 0$) become nearly uncorrelated.
{\bf (b, d)} Lines are AFQMC results at $T/t = 0.33$ (solid) and $T/t = 0.50$ (dashed);
\ding{73}: BCS expectation value at $T=0$, $\times$: from AFQMC at $T/t = 0.1$, both at $n = 0.6$.}
\label{fig:ConditionalDensityFigure}
\end{figure*}

The collection of atom-resolved data gives access to thermodynamic quantities as well as spin and charge correlations~\cite{cheuk2016observation,hartke2020doublon}. We use these correlations to establish the evolution of the spin-imbalanced Hubbard gas from a mixture of pairs and excess fermions to a Fermi liquid of polarons.
As the most direct measure of pair correlations, we show the doublon density $d$ as a function of magnetization and density for various interaction strengths in Fig.~\ref{fig:PairingFigure}(a). The natural trend is an increase of $d$ with attraction strength or density and a decrease with magnetization, as the gas contains fewer minority atoms. Remarkably, already at moderate spin-doping the doublon density closely approaches the value one finds assuming each minority atom forms a Fermi polaron (dotted lines)~\cite{Chevy2006, Pascual2024, SI}. To bring out this limit, the doublon density per minority atom $d/n_\downarrow$ is of interest or, for better visibility, its complement: the fraction of singlons among minority atoms $s_\downarrow/n_\downarrow$, which we show in Fig.~\ref{fig:PairingFigure}(b).
While increasing attraction naturally suppresses singlons, a significant singlon fraction is expected both for nonlocal fermion pairs as well as for Fermi polarons, given the finite size of their dressing cloud.
Indeed, the data at large magnetization approaches the singlon fraction expected for a Fermi polaron from the variational Ansatz~\cite{Chevy2006, Pascual2024, SI}.
This provides evidence that the gas forms a Fermi liquid of polarons in this limit.

Fluctuations and correlations of our atom-resolved data provide further insight into the nature of fermion pairs and polarons in the system.
In spin-balanced gases, full fermion pairing implies a vanishing spin susceptibility~\cite{Hartke2022Direct}. We obtain the spin susceptibility $\chi_m$ directly from the measured total spin fluctuations via the fluctuation-dissipation theorem: $\chi_m T=\sum_\delta \langle \hat{m}_i\hat{m}_{i+\delta} \rangle_c$~\cite{hartke2020doublon, Hartke2022Direct, zhou2011universal}. 
At strong interaction ($U/t \gtrsim 6$) and spin balance, the observed total spin fluctuation indeed vanishes~(Fig.~\ref{fig:PairingFigure}(c)), implying full pairing~\cite{Hartke2022Direct}.

With spin imbalance, spin fluctuations are seen to increase.
Positive spin fluctuations are expected even in the idealized case where pairing is still complete, as the gas of unpaired excess fermions will fluctuate at finite temperature.
To determine whether minority atoms remain fully paired at finite magnetization, it suffices to show that they do not contribute to the total spin fluctuations.
Figure~\ref{fig:PairingFigure}(c) reveals that at $U/t \gtrsim 6$ and for small magnetization ($m\lesssim 0.05$) spin fluctuations are nearly independent of the density $n$ at constant magnetization, i.e. $\partial(\chi_m T)/\partial n |_m\approx0$. Physically, this implies that increasing the density at fixed $m$ by adding equal numbers of spin-up and spin-down atoms does not increase the spin fluctuations. These newly added atoms must therefore form pairs.
We conclude that for interaction strengths $U/t \gtrsim 6$, and small magnetization $m \lesssim 0.05$, all minority fermions remain paired, and the remaining magnetic fluctuations are due to residual fluctuations of excess fermions.
We note that the pair criterion of vanishing $\partial(\chi_m T)/\partial n |_m$ is thermodynamically equivalent to the flatness of the singlon density profiles ($\partial m/\partial \mu = 0$) observed in Fig.~\ref{fig:IntroFigure}(c)~\cite{SI}.
For larger magnetization $m \gtrsim 0.05$ and densities $n \gtrsim 0.4$ the nearly flat magnetization profile in Fig.~\ref{fig:IntroFigure}(c) suggests that the gas still features strong pairing correlations, but the increasing dependence of the spin susceptibility on density in Fig.~\ref{fig:PairingFigure}(c) indicates the evolution from a mixture of pairs and excess fermions towards the regime of a Fermi liquid of polarons.
We do not observe a sudden transition in the evolution from pairs to polarons, presumably due to our non-zero temperatures and because the Fermi polaron's quasiparticle weight becomes small at strong interactions, causing it to closely resemble a molecule~\cite{Schirotzek2009, kohstall2012metastability, ness2020observation, Pascual2024}.

Having demonstrated that minority atoms remain fully paired at small magnetization while evolving into Fermi polarons at large magnetization, we now investigate the spatial structure of pair correlations. The interplay of attraction between unlike spins, Pauli exclusion between like spins, and repulsion between fermion pairs leads to complex nonlocal correlations.
Interspin density correlations are dominated by charge-density wave (CDW) correlations between doublons~\cite{Hartke2022Direct}. To elucidate the spatial structure of pairs, we instead focus on the correlations between isolated spins (singlons).
At spin balance, attraction causes unlike isolated spins to bunch~(Fig.~\ref{fig:ConditionalDensityFigure}(a)), reflecting the nonlocal wavefunction of fermion pairs, while like spins remain largely uncorrelated~(Fig.~\ref{fig:ConditionalDensityFigure}(c), left panel).
With spin doping, a degenerate Fermi gas of unpaired excess spins emerges.
The Fermi hole reflecting Pauli exclusion~\cite{cheuk2016observation} is directly revealed by negative nonlocal correlations~(Fig.~\ref{fig:ConditionalDensityFigure}(c), right panel).

To disentangle the precise way in which spin imbalance affects nonlocal fermion pairing, we measure the {\it excess} density of opposite spins around an isolated spin, defined as $\Delta s_\uparrow = \sum_\delta P(s_{i+\delta,\uparrow} | s_{i,\downarrow}) - P(s_{i,\uparrow})$, where $P(s_{i+\delta,\uparrow}|s_{i,\downarrow})$ is the probability of detecting an isolated spin-up atom at site-$(i+\delta)$ {\em conditioned on} an isolated spin-down atom on site-$i$, and  $P(s_{i,\uparrow})$ is the averaged density of isolated spin-up atoms. 
This quantity can be obtained from the intra-spin connected correlation and the density of spin-down atoms via $\Delta s_\uparrow = \sum_\delta \langle \hat{s}_{i+\delta,\uparrow} \hat{s}_{i,\downarrow}\rangle_c / \langle \hat{s}_{\downarrow} \rangle$. 

In the molecular limit, where the pair size is much smaller than the interparticle spacing ($U \gg t, n \ll 1)$, one additional spin-up atom around any isolated spin-down is expected. 
Indeed, this behavior is observed as the excess density approaches unity at positive polarization in Fig.~\ref{fig:ConditionalDensityFigure}(b) at $U/t=8.44$~(right panel). 
When the attraction is reduced, the size of the pairs increases and becomes comparable to the interatomic spacing~\cite{He2022}. At $U/t=5.76$ and spin balance ($m=0$) in Fig.~\ref{fig:ConditionalDensityFigure}(b), we measure an excess density less than one.
In this form of pairing, each single atom attracts only a portion of an opposite spin while being subject to strong Pauli exclusion with alike spins, resulting in a fully paired gas of vanishing spin susceptibility~(implied by Fig.~\ref{fig:PairingFigure}(b) and Fig.~\ref{fig:IntroFigure}(d)).

\begin{figure*}[!t]
	\centering
	\includegraphics[width=1.6\columnwidth]{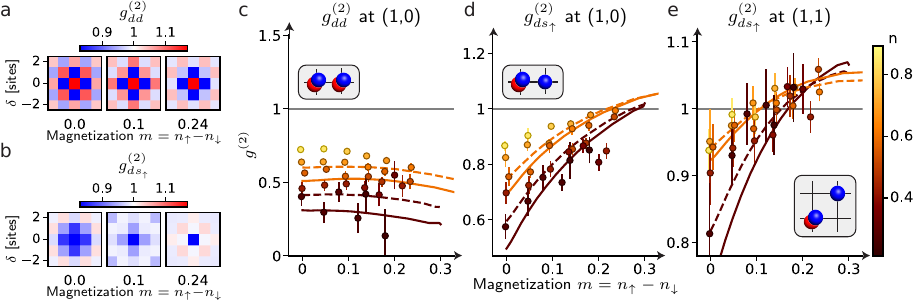}
	\caption{\textbf{Effective pair-pair and pair-dopant interactions.} 
{\bf (a)}~Measured doublon-doublon correlations ($g^{(2)}_{d,d} = \langle \hat{d}_i \hat{d}_{i+1}\rangle/d^2$) at three different magnetizations and $n\approx0.7$ showing pair-pair repulsion.
{\bf (b)}~Pair-dopant repulsion, detected through doublon-singlon up correlations ($g^{(2)}_{d,s_\uparrow} = \langle \hat{d}_i \hat{s}_{\uparrow,i+1}\rangle/d s_\uparrow$) at $n\approx0.7$. 
{\bf (c)}~
Doublon-doublon repulsion is dependent only on density, revealed by $g^{(2)}_{d,d}$ at displacement $(1,0)$.
{\bf (d-e)}~In contrast, at high magnetization and high density, doublon-singlon repulsion is overcome by many-body effects, producing positive correlations in $g^{(2)}_{d,s_\uparrow}$ at nearest-neighbor~(d) and next-nearest-neighbor~(e) displacements.
Data is shown for $U/t=5.76$.
Lines in (c-e) are AFQMC results at $T/t = 0.33$ (solid) and $T/t = 0.50$ (dashed).
	}
	\label{fig:BosonFigure}
\end{figure*}

\begin{figure*}[!t]
	\centering
	\includegraphics[width=7 in]{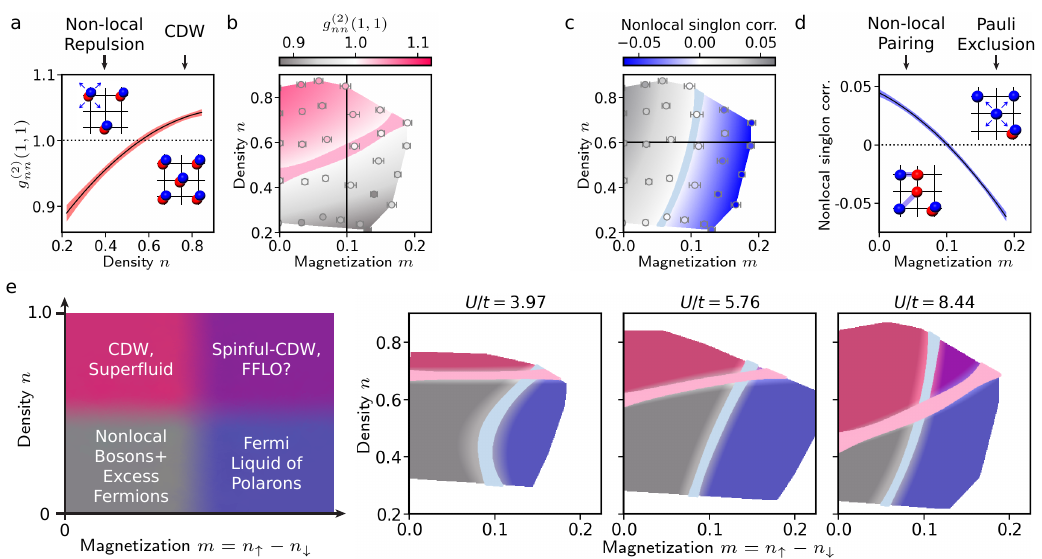}
	\caption{\textbf{Classifying emergent behavior of a spin-imbalanced Hubbard gas with microscopic observables.} 
{\bf (a)}~Many-body order is revealed by the sign of the diagonal density-density correlation ($g^{(2)}_{n,n}(1,1) = \langle \hat{n}_i \hat{n}_{i+(1,1)}\rangle/n^2$). At low density fermion pairs become larger and repel, leading to negative diagonal density correlations. At high density, pairs order into a charge-density wave~(CDW) having positive diagonal density correlations. 
{\bf (b)}~Two-dimensional map of $g^{(2)}_{n,n}(1,1)$ measured at various densities and magnetizations with $U/t=8.44$. Markers show experimental data, and background color shows a second order two dimensional polynomial interpolation. The pink band is the crossover region where $g^{(2)}_{n,n}(1,1)\approx 1$ within
the statistical uncertainty of the fit (1$\sigma$ confidence level from bootstrapping). Data in (a)~shows a cut of the interpolation in (b) at $m=0.1$, including statistical uncertainty~(red shading).
{\bf (c-d)}~The dominant behavior of isolated fermions is revealed by the sign of the total nonlocal singlon-singlon~(moment-moment) correlations. Isolated fermions include both excess majority fermions and the nonlocal fluctuations of bosonic pairs. Total nonlocal singlon correlations cross over from pairing dominated~(positive) at low magnetization to Fermi dominated~(negative) at high magnetization. Data in (c-d) is analogous to (a-b), and (d) shows a cut through (c) at $n=0.6$. 
{\bf (e)}~Resulting experimental classification of the emergent order of the spin-imbalanced mixture for multiple attraction strengths, ranging from strongly attractive~($U/t=8.44$) to below the threshold for full pairing~($U/t=3.97$). Each map is obtained using the the observables in (a-d), with similar statistical certainty.
	}
\label{fig:PhaseDiagramFigure}
\end{figure*}

This ``partial pairing'' is the real-space manifestation of pairing constrained by Pauli blocking. BCS theory provides an intuitive picture: only a fraction $\Delta/E_F$ of fermions near the Fermi surface participate in the pairing, where $\Delta$ and $E_F$ are the pairing gap and Fermi energy, respectively.
Indeed, when low temperature ($T/t=0.1$) AFQMC simulations and analytical BCS theory calculations are compared for $\Delta s_\uparrow$ (Fig.~\ref{fig:ConditionalDensityFigure}(b)), the results agree strikingly well for any interaction strength, showing that BCS theory can capture this aspect of partial pairing well in the Hubbard gas.
The agreement between the data at finite temperature and the zero temperature expectation becomes better with increasing attraction $U/t$. This is natural as the pairing gap increases with $U/t$, suppressing temperature effects.

The excess majority spin density ($\Delta s_\uparrow$ for $m>0$) does not depend significantly on $m$ throughout the crossover, suggesting that minority atoms attract a similar portion of a majority atom in the pairing and polaron regimes for intermediate and large $U/t$.

The non-local correlations of majority excess singlons in Fig.~\ref{fig:ConditionalDensityFigure}(d) clearly display a Fermi hole. Its magnitude depends only on the number $m=n_\uparrow-n_\downarrow$ of excess fermions, but not on the total density $n$ of the gas, even when pairs are significantly nonlocal ($U/t = 5.76$).
This is the natural expectation for a Bose-Fermi mixture, where only excess fermions form a Fermi sea. But even in the polaron Fermi liquid with two Fermi seas of up and down spins, strong interactions cause about one majority atom for each minority spin to become part of the polarons' dressing cloud, leaving a reduced Fermi hole of effectively $m=n_\uparrow - n_\downarrow$ majority fermions.


After seeing that pairing correlations persist upon doping the system with degenerate excess spins, we turn to investigate how the interactions between pairs and dopants evolve as magnetization increases.
For a large portion of the parameter space, the gas can be viewed as an interacting mixture of pairs and dopants.
Pauli exclusion produces effective hard-core onsite repulsion between doublons and singlons. The same effect also results in repulsive nearest-neighbor doublon-doublon and doublon-singlon interactions by preventing quantum fluctuations of local doublons into nonlocal pairs.

The pair-pair repulsion manifests as strong antibunching of doublons at adjacent sites~($g^{(2)}_{d,d}(1)=\langle \hat{d}_i \hat{d}_{i+(1,0)}\rangle/\langle\hat{d}\rangle^2$) and, for high enough doublon densities, positive correlation on the diagonal (Fig.~\ref{fig:BosonFigure}(a)), a hallmark of the charge-density wave (CDW) state~\cite{Hartke2022Direct}. 
The nearest neighbor doublon-doublon repulsion decreases approaching half filling, as the Fermi hole shrinks in size, but it is largely independendent of magnetization (Fig.~\ref{fig:BosonFigure}(c)).
In contrast, doublons and majority singlons repel only at low magnetization (Fig.~\ref{fig:BosonFigure}(b,d-e)). Remarkably, we observe a transition from repulsive $d-s_\uparrow$ correlations at low magnetization to weak attraction at high magnetization. Such bunching of doublons and majority singlons cannot be explained from the variational Ansatz for a single Fermi polaron, which always predicts negative correlations arising from Fermi repulsion among majority fermions~\cite{Chevy2006, Pascual2024, SI}.
The bunching beyond a particular magnetization instead indicates a many-body effect. When doublons and singlons are in close proximity, the minority atom in the doublon pair can delocalize between the two sites without an interaction energy penalty---reducing its kinetic energy.
The non-local attraction of singlons and doublons indicates the system's tendency to form a spin-charge density wave. Such simultaneous ordering of pairs and dopants in the gas may serve as a precursor of FFLO-type superfluidity at lower temperatures~\cite{he2006finite, feng2025search}.


With the help of the spatially resolved spin and charge correlations, we characterize the competition of different orders within the spin-imbalanced Hubbard gas, resulting in a ``phase diagram'' of correlated behavior. We first investigate the many-body ordering of the doublon pairs with changing density. Pauli exclusion driven repulsion governs the dynamics in the low density regime, so the cloud can be understood as a dilute repulsive bosonic gas (left side of Fig.~\ref{fig:PhaseDiagramFigure}(a)).
In contrast, the competition between the nearest neighbor interaction and kinetic energy at higher density is in favor of a charge-density wave that can be identified by a positive diagonal correlation~(right side of Fig.~\ref{fig:PhaseDiagramFigure}(a)).
The crossover from negative to positive density-density correlation indicates a transition from a dilute repulsive doublon gas to a many-body ordered state. The crossover point varies with both density and magnetization. At $U/t=8.44$~(Fig.~\ref{fig:PhaseDiagramFigure}(b)), the crossing~(red band) occurs at higher density when the gas is spin doped. The observed trend is in agreement with NLCE calculations at an elevated temperature of $T/t=0.8$ (Fig.~\ref{fig:nlce}(a)) and can be understood as the spin doping reducing the number of the bosonic pairs at a constant density. 

Similarly, we can characterize the presence of an effective degenerate Fermi gas of the excess fermions by looking at the total nonlocal singlon-singlon correlations (regardless of spin) defined as $\sum_{\delta {\neq} 0} \langle s_i s_{i+\delta}\rangle_c$, shown in Fig.~\ref{fig:PhaseDiagramFigure}(c-d) (NLCE theory results shown in Fig.~\ref{fig:nlce}(b)). This metric is dominated by the presence of nonlocal pairs at low magnetization, resulting in a positive net correlation. With increased magnetization, the metric gradually changes sign due to unpaired, excess fermions forming a degenerate Fermi gas and experiencing Fermi pressure. The crossing point from positive to negative correlations is primarily independent of density. The residual dependence on density in Fig.~\ref{fig:PhaseDiagramFigure}(c) indicates an enhanced pairing strength close to half-filling. 

In Fig.~\ref{fig:PhaseDiagramFigure}(e), we combine the measurements of many-body ordering in Fig.~\ref{fig:PhaseDiagramFigure}(a-b) with the measurements of the emerging Fermi sea in Fig.~\ref{fig:PhaseDiagramFigure}(c-d) to form a ``phase diagram'' of correlations. Furthermore, we extend these measurements to three other interaction strengths, classifying the observed emergent behavior for varying $m$, $n$, and $U/t$, all near $T/t \sim 0.3$. 
The phases and boundaries shown are derived through the procedure in Fig.~\ref{fig:PhaseDiagramFigure}(b) and Fig.~\ref{fig:PhaseDiagramFigure}(c), with the thickness indicating the statistical uncertainty for the crossover. Above the pink line, CDW correlations exist. To the right of the blue line, the Fermi sea has emerged. 
The purple region of the phase space featuring both many-body ordering and a Fermi sea is a promising candidate for a potential FFLO phase at lower temperatures.


In conclusion, we realize the spin and charge doped Hubbard model with fermionic $^{40}$K at single atom resolution, finding a crossover from a mixture of fermion pairs and excess dopants to a Fermi liquid of polarons. 
Pairs are nonlocal and overlapping, leading to ``partial pairing''.
While our static observables do not display a clear transition in behavior from fermion pairs to polarons, future studies of the system's dynamical response may reveal the presence or absence of quasiparticle weight of minority spins~\cite{Schirotzek2009, koschorreck2012attractive, parish2021thermodynamic, Ji_2021, prichard2025magnon}.
The correlation diagram reveals an intriguing regime of simultaneous pair-pair repulsion and pair-dopant bunching, indicating the formation of a spin-charge-density wave. This many-body order of fermion pairs coexisting with excess dopants may provide a precursor to FFLO superfluidity~\cite{feng2025search}.

This work was supported by the NSF through the Center for Ultracold Atoms,
PHY-2012110 and PHY-2513210, AFOSR (FA9550-23-1-0402), DOE (DE-SC0024622), and the Vannevar Bush Faculty Fellowship (ONR N00014-19-1-2631).
We thank the Flatiron Institute Scientific Computing Center for computational resources. The Flatiron Institute is a division of the Simons Foundation.
C.T. acknowledges the support from the National Science Foundation Graduate Research Fellowship Program (NSF GRFP) under Grant No. 2141064.
E.K. was supported by the NSF under Grant No. DMR-1918572.
J.H. gratefully acknowledges support by the MIT Pappalardo Fellowships in Physics.
Computations for NLCE were performed on Spartan high-performance computing facility at San Jos\'{e} State University, which was supported by the NSF under Grant No. OAC-1626645.

\bibliographystyle{apsrev4-2} 
\bibliography{SpinImbalance}
\clearpage 

\setcounter{equation}{0}
\setcounter{figure}{0}
\setcounter{secnumdepth}{2}
\renewcommand{\theequation}{S\arabic{equation}}
\renewcommand{\thefigure}{S\arabic{figure}}
\renewcommand{\tocname}{Supplementary Materials}
\renewcommand{\appendixname}{Supplement}


\onecolumngrid

\section*{Supplementary Information}

\subsection{Experimental procedure}
We realize the spin-imbalanced attractive Hubbard gas using the two lowest energy states of $^{40}$K as described in previous work~\cite{Hartke2022Direct, hartke2020doublon, cheuk2015quantum}.
The gas is trapped in a single two dimensional plane of a three dimensional optical lattice which also provides the underlying harmonic confinement.
The two dimensional lattice potential spacing is $a_x \approx a_y \approx \SI{541}{nm}$ and the potential depth is measured to be $4.3(2) E_R$ yielding a tunneling energy $t = h \times \SI{340 \pm 20}{Hz}$. Here $E_R = \hbar^2 \pi^2 / (2 m_K a_{x,y}^2) = h \times \SI{4260}{Hz}$ is the recoil energy, ($\hbar$) $h$ is the (reduced) Planck constant, and $m_K$ is the mass of a single $^{40}$K atom.
The in-plane harmonic trapping frequency provided by the envelope of the optical lattice potential is $\omega_{x,y} = 2 \pi \times \SI{26 \pm 0.9}{Hz}$. The out-of-plane lattice spacing is $a_z \approx 3 \mu m$ and the trapping frequency provided by the optical lattice is $\omega_{z} = 2 \pi \times \SI{4.5 \pm 0.1}{kHz}$. The interaction between different spin species is controlled using the ambient magnetic field near an $s$-wave Feshbach resonance at $\SI{202.1}{G}$ and we use the numerically calculated wavefunctions to calculate the attraction energy $U$ from the scattering length.

To prepare a spin-imbalanced mixture, we start with a spin-polarized gas in the lowest hyperfine state of $^{40}$K, and drive an imperfect Landau-Zener radiofrequency (RF) sweep to the neighboring hyperfine state.
Tuning the strength of the RF coupling changes the spin imbalance after the sweep but before evaporative cooling.
After each experimental realization of the system the atoms' position is suddenly frozen by increasing the lattice depth to $\sim 100 E_R$. With the help of Stern-Gerlach separation into a bilayer potential, spin-selective, single-atom-resolved images of each spin state are obtained, yielding the complete spin and charge state at every lattice site.
Post-selecting the experimental snapshots based on magnetization gives access to measurements at a particular magnetization value.

\subsection{Relation of pairing and a flat polarization profile}
Here we show how pairing follows directly from a flat magnetization profile.
For a balanced gas, complete pairing is implied by the vanishing of total spin fluctuations, and thus (at finite temperature) a zero spin susceptibility $\chi_m$.
However, for non-zero magnetization, $\chi_m$ will be non-zero due to the contribution from unpaired excess fermions, even if all minority atoms are still fully paired. A criterion for complete pairing that holds for non-zero magnetization is then given by
\begin{equation}\label{eq:si_flat_profile_pairingcondition}
\left. \frac{\partial \chi_m}{\partial n}\right|_m = 0.  
\end{equation}
Indeed, if the addition of particles at constant magnetization does not increase the magnetic susceptibility, then those particles have come in pairs.

We can now relate this criterion to the observation of a flat polarization profile in the trap, which implies
\begin{equation}\label{eq:si_flat_profile}
    \left. \frac{\partial m}{\partial \mu}\right|_h = 0
\end{equation}
in an extended region of values $\mu$ and $h$. Here, $m(\mu, h)$ is the magnetization written as a function of the chemical potential and effective magnetic field $h$.
The trapping potential sweeps the chemical potential radially across the cloud in the local density approximation but it is keeping $h$ constant.
From the flat polarization, we deduce that the spin susceptibility is also not changing in the trap:
\begin{equation}\label{eq:si_flat_profile_derivative}
\left. \frac{\partial \chi_m}{ \partial \mu} \right|_h = \left. \frac{\left. \frac{\partial m}{\partial h}\right|_\mu}{\partial \mu} \right|_h = \left. \frac{\left. \frac{\partial m}{\partial \mu}\right|_h}{\partial h} \right|_\mu = 0.
\end{equation}

To see whether the condition of complete pairing Eq.~\ref{eq:si_flat_profile_pairingcondition} is implied by the flat polarization profile Eq.~\ref{eq:si_flat_profile}, we find the change of magnetization with parameters $\mathrm{d} \mu$ and $\mathrm{d} h$:
\begin{equation}
    \mathrm{d}m = \left. \frac{\partial m}{\partial \mu}\right|_h \mathrm{d}\mu + \left. \frac{\partial m}{\partial h}\right|_\mu \mathrm{d}h
\end{equation}
As the pairing criterion demands constant magnetization $\mathrm{d}m = 0$, and since $\chi_m = \partial m / \partial h |_\mu$ is non-zero for $m>0$, the flat polarization condition implies that then also $\mathrm{d}h = 0$.

Under this condition, the change in the spin susceptibility is
\begin{equation}
    \mathrm{d} \chi_m = 
    \left. \frac{\partial \chi_m}{ \partial \mu} \right|_h \mathrm{d}\mu 
    + \left. \frac{\partial \chi_m}{ \partial h}\right|_\mu \mathrm{d}h = 0
\end{equation}
where we used Eq.~\eqref{eq:si_flat_profile_derivative} and $\mathrm{d}h = 0$.
Therefore indeed, a flat polarization profile implies complete pairing, expressed in the condition $\partial \chi_m/ \partial n|_m=0$.

\subsection{Theoretical methods}
\label{SI:Numerical}

\subsubsection{The Fermi polaron in the attractive Hubbard model}
The Fermi polaron in the attractive Hubbard model was found in~\cite{Pascual2024} to be well described by the variational Ansatz~\cite{Chevy2006}:
\begin{equation}
    \ket{\psi} = \alpha_0 c^\dagger_{0,\downarrow} \ket{\text{FS}_\uparrow}
    + \sum\limits_{k,q} \alpha_{k,q} c^\dagger_{q-k,\downarrow} c^\dagger_{k,\uparrow} c_{q,\uparrow} \ket{\text{FS}_\uparrow}
\end{equation}
where $c^\dagger$ and $c$ are momentum state creation and annihilation operators of the different spin states and $\ket{\text{FS}_\uparrow}$ is the Fermi sea of the spin-up fermions.
The momentum sum is over $k>k_F$ and $q \leq k_F$ where $k_F$ is the Fermi momentum.
The wavefunction amplitudes $\alpha_0$ and $\alpha_{k,q}$ can be found by minimizing the energy expectation value~\cite{Pascual2024}.
The quasiparticle weight is $Z_0 = \alpha_0^2$.
The doublon fraction of the polaron $d/n_\downarrow$ can be found either by calculating the doublon expectation value of the wavefunction directly or by taking the derivative of the polaron energy with attraction strength $U$.
The polaron wavefunction describes a single spin down fermion in a sea of spin up fermions, where the magnetization equals the density ($m=n$).
For smaller magnetizations in Fig.~\ref{fig:PairingFigure}(a) the expected doublon density was calculated assuming a constant polaron doublon fraction $d/n_\downarrow$ given by the polaron wavefunction.
The wavefunction also gives access to other correlators, like the doublon-singlon-up correlators.

\subsubsection{Quantum Monte Carlo}
We employ state-of-the-art quantum Monte Carlo computations in this work. A combination of the determinant quantum Monte Carlo (DQMC) method \cite{BSS} and finite-temperature constrained-path (CP) auxiliary-field quantum Monte Carlo (AFQMC) method \cite{Zhang1999, He2019} is used to simulate the spin-imbalanced attractive Hubbard model. To evaluate the partition function, we apply the Trotter decomposition to break the inverse temperature into discrete imaginary time slices: $\beta = \Delta\tau \times L_{\tau}$. The on-site interaction term is decoupled using the Hubbard-Stratonovich transformation, which introduces auxiliary fields and maps the many-body problem onto a sum over one-body problems. Configurations of the auxiliary fields are sampled via a Monte Carlo procedure.

At the higher-temperatures we studied,
the order of magnitude of the average sign for the parameter regimes is 0.1. We
ensured sufficient statistical accuracy by performing a sufficiently large number of warm-up and measurement sweeps. For high temperatures cross-checks with CP AFQMC are performed. We have also verified via cross checks and comparisons of low-temperature results against ground-state AFQMC results that any effect from the so-called infinite variance problem \cite{Shi2016} is minimal in our DQMC results. A time step of $\Delta\tau = 0.05$ is used in this work, where the Trotter error is negligible compared to statistical errors. All simulations are performed on a $16 \times 16$ square lattice with periodic boundary conditions. We have verified that finite-size effects are limited, particularly for the near-neighbor correlations presented in this paper.

\subsubsection{Numerical Linked-Cluster Expansions}

The numerical linked-cluster expansion (NLCE)~\cite{M_rigol_06,b_tang_13b} expresses the correlation functions and other extensive properties of a lattice model as a series written directly in the thermodynamic limit (no systematic finite-size errors are present). The expansion is in terms of contributions from all finite clusters that can be embedded in the lattice. These contributions are calculated exactly using full diagonalization of the Hamiltonian, which in turn sets the limit for the largest clusters that can be considered in the series. Here, we adopt a site expansion for the square lattice, namely, in order $l$, we take into account contributions from all clusters with up to $l$ sites, and carry out the expansion to order 9 for the Hubbard model. Since exact diagonalization is at the core of computing observables for the model on finite clusters, they are readily available on the entire grid of chemical potential, magnetic field, and temperature in a single run. We use numerical resummations using the Euler and Wynn algorithms~\cite{b_tang_13b} to extend the convergence of the series to lower temperatures. We find that $T/t=0.8$ is typically the lowest convergence temperature for properties measured in the experiment and that NLCE results agree with those obtained from DQMC and CPQMC methods at the same temperature.

\begin{figure}[tbhp]
    \includegraphics[width=0.5\linewidth]{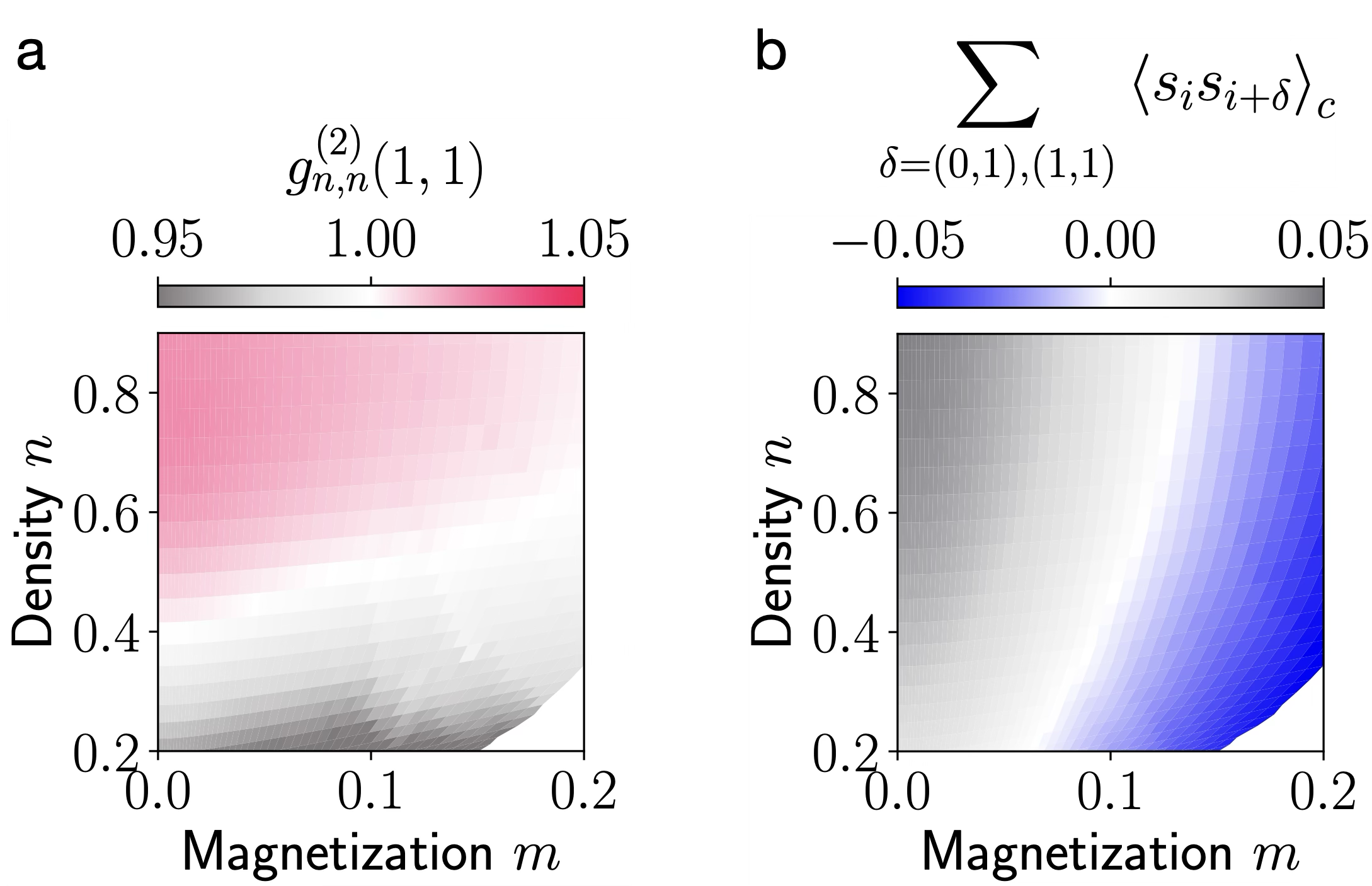}
    \caption{NLCE results for {\bf (a)} the diagonal density-density correlation $g^{(2)}_{n,n}(1,1)$, and {\bf (b)} the nonlocal singlon-singlon~(moment-moment) correlations, here the sum of the correlations between nearest and diagonal neighbors, in the space of the density and the magnetization for $U/t=8.4$ and at $T/t=0.8$.}
    \label{fig:nlce}
\end{figure}

In Fig.~\ref{fig:nlce}, we show NLCE results for the diagonal density-density correlation $g^{(2)}_{n,n}(1,1)$ and the nonlocal singlon-singlon correlations in the space of the density and the magnetization for $U/t=8.4$ and at $T/t=0.8$. Although the results are at an elevated temperature, they display a very good qualitative agreement with those obtained in the experiment, shown in Figs.~\ref{fig:PhaseDiagramFigure}(b) and \ref{fig:PhaseDiagramFigure}(c) of the main text, and point to very similar crossover regions.

\end{document}